\documentclass[journal]{IEEEtran}
\ifCLASSINFOpdf
   \usepackage[pdftex]{graphicx}
   \graphicspath{{../pdf/}{../jpeg/}}
   \DeclareGraphicsExtensions{.pdf,.jpeg,.png}
\else
\fi
\usepackage{mathtools}
 \usepackage{booktabs}
\usepackage{multirow}
\usepackage{algorithm}
\usepackage{algpseudocode}
\usepackage{amsmath}
\usepackage{graphicx}
\usepackage{bbm}
\usepackage{amssymb}
\usepackage[inline]{enumitem}

\usepackage[hidelinks]{hyperref}
\hypersetup{hidelinks}
\algnewcommand{\LeftComment}[1]{\Statex \(\triangleright\) #1}
\usepackage{tikz}
\usetikzlibrary{shapes,arrows,shadows,positioning}
\usepackage{amsmath,bm,times}
\usepackage{verbatim}
\usepackage{cite}
\begin{document}

\pgfdeclarelayer{background}
\pgfdeclarelayer{foreground}
\pgfsetlayers{background,main,foreground}

\tikzstyle{block}=[draw, fill=white!20, text width=7em, 
    text centered, minimum height=2.5em,drop shadow,rounded corners]
\tikzstyle{block2}=[draw, text width=5em, 
    text centered, minimum height=1.5em,rounded corners,fill=white!20]
\tikzstyle{process}=[draw, fill=white!20, text width=5em, text centered, minimum height=2.5em,drop shadow]
\tikzstyle{process2}=[draw, fill=white!20, text width=3em, text centered, minimum height=2.5em,drop shadow]
\tikzstyle{qtion}=[shape aspect=2, diamond, draw, fill=white!20, text width=5em, 
    text centered, minimum height=1.5em,drop shadow]
    \tikzstyle{ball} = [draw, fill=black!100, circle, radius=1mm,scale=0.6]
\tikzstyle{ball_2} = [ultra thin]


\tikzstyle{materia}=[draw, thick, text width=30em, text centered,
minimum height=1.5em, drop shadow]
\tikzstyle{practica} = [materia, fill=white!20, text width=30em, minimum width=30em,
minimum height=3em, rounded corners, drop shadow]
\tikzstyle{texto} = [above, text width=20em]
\tikzstyle{linepart} = [draw, thick, -latex', dashed]
\tikzstyle{line} = [draw, thick, -latex']
\tikzstyle{ur}=[draw, text centered, minimum height=0.01em]

\newcommand{\blockdist}{1.3}
\newcommand{\edgedist}{1.5}

\newcommand{\practica}[2]{node (p#1) [practica]
{\\{\normalsize\textrm{#2}}}}


\newcommand{\background}[5]{%
\begin{pgfonlayer}{background}
\path (#1.west |- #2.north)+(-0.6,0.6) node (a1) {};
\path (#3.east |- #4.south)+(+0.6,-0.6) node (a2) {};
\path[fill=gray!20, rounded corners, draw, thick, dashed]
(a1) rectangle (a2);
\path (a1.east |- a1.south)+(3.6,-0.6) node (u1)[texto]
{\large{\textit{\textbf{Module #5}}}};
\end{pgfonlayer}}

\newcommand{\transreceptor}[3]{%
\path [linepart] (#1.east) -- node [above]
{\scriptsize Transreceptor #2} (#3);}

%
\title{Decentralized P2P Energy Trading under Network Constraints in a Low-Voltage Network}
%
%
%
\author{Jaysson~Guerrero\textsuperscript{\href{https://orcid.org/0000-0001-6945-0307}{\includegraphics[scale=0.07]{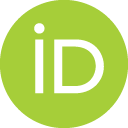}}},~\IEEEmembership{Student Member,~IEEE,}        Archie~C.~Chapman\textsuperscript{\href{https://orcid.org/0000-0002-5055-3004}{\includegraphics[scale=0.07]{orcid.png}}},~\IEEEmembership{Member,~IEEE,} and~Gregor~Verbi\v{c}\textsuperscript{\href{https://orcid.org/0000-0003-4949-768X}{\includegraphics[scale=0.07]{orcid.png}}},~\IEEEmembership{Senior~Member,~IEEE}
\thanks{The authors are with The University of Sydney, School of Electrical and Information Engineering, NSW, 2006, Australia (e-mail: jaysson.guerrero@sydney.edu.au; gregor.verbic@sydney.edu.au; archie.chapman@sydney.edu.au).}
}

\maketitle

\begin{abstract}
The increasing uptake of \textit{distributed energy resources} (DERs) in distribution systems and the rapid advance of technology have established new scenarios in the operation of low-voltage networks. In particular, recent trends in cryptocurrencies and blockchain have led to a proliferation of \textit{peer-to-peer (P2P) energy trading} schemes, which allow the exchange of energy between the neighbors without any intervention of a conventional intermediary in the transactions. Nevertheless, far too little attention has been paid to the technical constraints of the network under this scenario. A major challenge to implementing P2P energy trading is that of ensuring that network constraints are not violated during the energy exchange. This paper proposes a methodology based on sensitivity analysis to assess the impact of P2P transactions on the network and to guarantee an exchange of energy that does not violate network constraints. The proposed method is tested on a typical UK low-voltage network. The results show that our method ensures that energy is exchanged between users under the P2P scheme without violating the network constraints, and that users can still capture the economic benefits of the P2P architecture.
\end{abstract}


%
\IEEEpeerreviewmaketitle

\section*{Nomenclature}
\addcontentsline{toc}{section}{Nomenclature}
\begin{IEEEdescription}[\IEEEusemathlabelsep\IEEEsetlabelwidth{$V_1,V_2,1$}]
\item[$\mathcal{T}$] Set of all time-slots $t$.
\item[$\mathcal{H}$] Set of all households. 
\item[$\mathcal{B}$] Set of all buyers.
\item[$\mathcal{S}$] Set of all sellers.
\item[$\mathcal{N}$] Set of all nodes $i$ in the network. 
\item[$\mathcal{E}$] Set of distribution lines connecting the nodes in the network. 
\item[$x^{+/-}$] Electrical power flowing from/to grid.
\item[$s^{+/-}$] Import and export tariffs.
\item[$\pi_{b}$] Bid price of buyer $b$. 
\item[$\pi_{s}$] Ask price of seller $s$.
\item[$\sigma_{b}$] Quantity of energy to purchase by buyer $b$. 
\item[$\sigma_{s}$] Quantity of energy to supply by seller $s$.
\item[$C_{i}^{c}$] Marginal benefit of consumer $c$.
\item[$C_{i}^{p}$] Marginal cost of prosumer $p$.
\item[$P_{i}^{c}$] Real power consumption of consumer $c$.
\item[$P_{i}^{p}$] Real power generation of prosumer $p$.
\item[$L_{\text{min}}$] Minimum value of bidding offers. 
\item[$L_{\text{max}}$] Maximum value of bidding offers.

\item[$\Phi_{kl}^{ij}$] Power transfer distribution factor of line ($k,l$) due to changes in nodes $i$ and $j$.
\item[$\Psi_{kl}$] Injection shift factor of a line connecting nodes $k$ and $l$.
\item[$P_{loss}$] Active power losses.
\item[$\text{BEC}^{ij}$] Bilateral exchange coefficient due to a bilateral transaction between nodes $i$ and $j$.

\end{IEEEdescription}

\section{Introduction}
%
%
%
%
\IEEEPARstart{T}{he} role of distributed energy resources (DERs) characterizes the future of electrical power systems. Photovoltaic (PV) panels, battery storage systems, smart appliances and electric vehicles are some of the resources that allow traditional domestic consumers to become \textit{prosumers}. In fact, end-users can already undertake control actions to manage their consumption and generation. This context has introduced new opportunities and challenges to power systems.
Local energy trading between consumers and prosumers is one of the new scenarios of growing importance in the domain of distribution networks. Local distribution markets have been proposed as means of efficiently managing the uptake of DERs \cite{AEMC,PINSON}. This involves the creation of new roles and market platforms that allow the active participation of end-users and the direct interaction between them. This scenario brings potential benefits for the grid and users, by facilitating: 
\begin {enumerate*} [label=(\roman*\upshape)]
\item the efficient use of demand-side resources, \item the local balance of supply and demand, as well as \item opportunities for users to receive economic benefits through sharing and using clean and local energy.
\end {enumerate*}

Given this context, a decentralized peer-to-peer (P2P) architecture has been proposed to implement local energy trading. Unlike to the traditional scheme, under a P2P scheme, prosumers can trade their energy surplus with neighboring users. Currently, the implementation of decentralized market platforms is possible due to new advances in information and communication technology, such as \textit{blockchain} and other \textit{distributed ledger technologies} (DLTs), which support transparent and decentralized transactions. Many studies have already considered DLTs as the base of their P2P energy trading platforms \cite{blockchain_microgrid,APP_BC}. 
For example, \cite{ EV_P2P} proposed a P2P energy trading model for electrical vehicles, showing the potential of blockchain to enhance cybersecurity on the P2P transactions. Similarly, the work in \cite{munsing_blockchains_2017} demonstrates the benefits of a blockchain-based microgrid energy market using smart contracts. Additionally, commercial P2P trading pilots’ projects have also been implemented recently. 
Most of these create a cryptocurrency that is used to trade energy between users\footnote{Examples of DLTs in P2P energy trading include PowerLedger (https://powerledger.io), Enosi (https://enosi.io) and LO3 Energy (https://lo3energy.com).}.

However, electricity exchange is different from any other exchange of goods. Residential users are part of an electricity network, which imposes hard technical constraints on the energy exchange. Completely decentralized energy trading, without any coordination, compromises the operation of the network within its technical limits. Therefore, physical network constraints must be included in energy trading models. 

Despite the importance of the technical constraints, so far they have attracted little attention. \textcolor{black}{The work in \cite{blockchain_microgrid} introduces the application of the blockchain technology for energy trading as well as for technical operation. Although the variation in power losses due to the energy exchanges is evaluated, the impacts of each transaction on voltage and network capacity issues are not considered. More recently, works like that of \cite{munsing_blockchains_2017} and \cite{PINSON_2} used decomposition techniques to solve an optimal power flow in a distributed fashion for P2P energy trading. In a similar context, an alternative approach to account for network constraints and attribution of network usage cost is proposed in \cite{MORSTYN}. Nevertheless, there are still some elements of debate such as the market framework, and how external cost due to the power exchange and network coupling constraints (from the AC power flow) can be associated with the transactions.} 

In response to this shortcoming, in this paper, we extend the existing P2P energy trading scheme by explicitly taking into account the underlying network constraints at the distribution level. All transactions have to be validated during the bidding process, based on the network condition. Moreover, each transaction will be charged with the extra costs associated with the physical energy exchanged (i.e. due to losses). To our knowledge, this is the first model that integrates decentralized P2P energy trading with network constraints. Previous research either only focused on the DLTs technologies or did not consider the network constraints.

In summary, the contributions of this paper are as follows: 
\begin{itemize}
	\item[$\bullet$] We illustrate the importance of including network constraints in the models of P2P trading to prevent voltage and capacity problems in the network;
    \item[$\bullet$] We propose a novel methodology based on sensitivity analysis to asses the impact of the transactions on the network and to internalize the external cost associated with the energy exchange;
    \item[$\bullet$] We present the benefits that P2P trading under network constraints may bring to power systems and end-users, by comparing our method with other strategies proposed to prevent upcoming LV network issues;
    \item[$\bullet$] We demonstrate a specific implementation of our methodology for P2P energy trading, comprising consumers and prosumers, which shows that our method is feasible and thereby appropriate for P2P energy trading schemes.
\end{itemize}

The paper progresses as follows: The next section introduces pertinent concepts from the implementation of P2P energy trading, and illustrates why network constraints must be considered. This is followed by a description of the methodology in Section III. Section IV summarizes the trading mechanism scheme that the case study of this paper builds on. Section V presents the model of the case study and simulation results, and Section VI concludes the paper.

\section{Preliminaries}
\begin{figure}[t]
	\centering
	\includegraphics[width=2.5in]{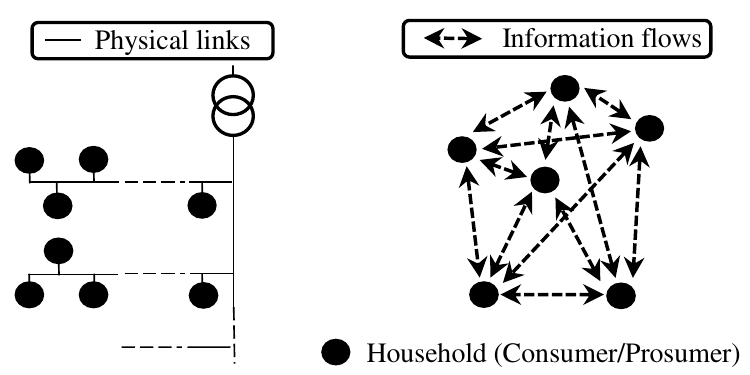}
	\caption{Model of information flows and physical links between households under a P2P scheme.}
	\label{P2P}
\end{figure}

Let $\mathbb{R}$ denote the set of real numbers, and $\mathbb{C}$ complex numbers. For a scalar, vector, or matrix $A$,  $A'$ denotes its transpose and $A^{*}$ its complex conjugate. The P2P scheme adopted is illustrated in Fig. \ref{P2P}. \textcolor{black}{The information flows between peers in a decentralized manner. As such, every peer can interact through financial flows with the others. It should be noted that the interaction channels (e.g. DLTs) are separate from the physical links}. The P2P scheme is composed of $H$ households agents, which are interacting among themselves over a decision horizon $\mathcal{T} \coloneqq \left \{ \tau, \tau + \Delta \tau,\ldots, \tau + T - \Delta \tau \right \}$ (typically one day) consisting of $T$ time-slots. Specifically, the network comprises a set of nodes $\mathcal{N} \coloneqq \left \{0,1, 2,\ldots, N \right \}$. We index the nodes in $\mathcal{N}$ by $i=0,1,\ldots,N$.

\subsection{Problem Description}

We consider a smart grid system for a P2P energy trading in a low-voltage (LV) network under a decentralized scheme. This paper considers the interaction of residential users through an online platform. Users can sell and buy energy to/from their neighbors or a retailer. We consider this a realistic assumption since currently there are pilot projects based on this concept, and it does not interfere with existing institutional arrangements (retail)\footnote{\textcolor{black}{Examples of pilot projects include Decentralized Energy Exchange (deX) Project, available at https://arena.gov.au/projects/decentralised-energy-exchange-dex/; and White Gum Valley energy sharing trial, available at https://westernpower.com.au/energy-solutions/projects-and-trials/white-gum-valley-energy-sharing-trial/.}}. A general P2P scheme is a method by which households interact directly with other households. Users are self-interested and have complete control of their energy used (different to centralized direct load control structures, in which some entity may have control of some appliances). 

Let $\mathcal{H}=\left \{ 1,2,\ldots,H\right \}$ be the set of all \textit{households} in the local grid. The time is divided into time slots $t \in \mathcal{T}$, where $\mathcal{T} = \left \{  1,2,\ldots,T\right \}$ and $T$ is the total number of time slots. The set of all households $\mathcal{H}$ is composed of the union of two sets: consumers $\mathcal{P}$ and prosumers $\mathcal{C}$ (i.e. $\mathcal{H} = \mathcal{P}\cup\mathcal{C}$). \textcolor{black}{We assume that all households are capable of predicting their levels of demand and generation for electrical energy for a particular time slot $t$. Specifically, we assume consumers bid in the market based on their demand profiles. As such, a demand profile is not divided into tasks or device utilization patterns, so that is the demand levels represent the total energy consumption over time}. Prosumers are classified into two types. Type 1 prosumers include those which have only PV systems; Type 2 includes prosumers which have PV systems, battery storage and \textit{home energy management systems} (HEMS). Prosumers have two options to sell their energy surplus:
\begin {enumerate*} [label=(\roman*\upshape)]
\item they can sell to the retailer and receive a payment for the amount of energy (e.g. feed-in tariff), or \item they can sell on the local market to consumers who participate in the P2P energy trading process. 
\end {enumerate*} 

\subsection{Household Agent Model}

A household $h \in \mathcal{H}$ uses $d_{t}^{h}$ units of electrical energy in slot $t$. Likewise, a household $h \in \mathcal{H}$ has $w_{t}^{h}$ units of energy surplus in slot $t$. The total quantity of electrical energy purchased in a slot $t$ is given by $x_{t}^{+}$, and its price is denoted by $s_{t}^{+}$. \textcolor{black}{The total energy consumption $x_{t}^{+}$ includes the amount of electrical energy purchased from the grid and from the local market}. Similarly, the quantity of electrical energy sold in a slot $t$ is given by $x_{t}^{-}$, and its price is denoted by $s_{t}^{-}$. While the energy surplus of Type 1 prosumers in $\mathcal{P}$ comes entirely from the PV system, each prosumer Type 2 in $\mathcal{P}$ uses its HEMS to optimize its self-consumption, considering their demand and energy surplus by solving the following mixed-integer linear programming (MILP) problem \cite{chanaka}:
\vspace{-0.001in}
\begin{align}\label{HEMS}
& \underset{x\in \mathcal{X}}{\text{minimize}} \quad \sum_{t \in \mathcal{T}}{(s_{t}^{+}x_{t}^{+} - s_{t}^{-}x_{t}^{-}})\\
&\text{s.t.} \quad \text{device operation constraints,} \nonumber\\ 
&\qquad \, \text{energy balance constraints, } \forall{t} \in \mathcal{T},\nonumber
\end{align}

\noindent where $\mathcal{X}$ is the set of decision variables $\left \{ x_{t}^{+},x_{t}^{-} \right \}$. State variables in the model are $s_{k}^{+} $ and $s_{t}^{-}$. The former is associated with the price of energy in time slot $t$, and the latter with the incentive received for the contribution to the grid. In other words, $s_{t}^{+} $ and $s_{t}^{-}$ are related to import tariffs (e.g. flat, time-of-use) or export tariffs (e.g. feed-in-tariff). The outcome of this process provides \textit{net load profiles} for users with HEMS. After their self-optimisation, prosumers can export their energy surplus to the grid.
\vspace{-0.005in}
\subsection{Network Model}
We consider a radial distribution network $\mathcal{G}(\mathcal{N}, \mathcal{E})$, consisting of a set of nodes $\mathcal{N}$ and a set of distribution lines (edges) $\mathcal{E}$ connecting these nodes. Using the notation of the branch flow model \cite{Nali2015}, we index the nodes by $i=0,1,\ldots,N$, where the root of our radial network (Node 0) represents the substation bus, and it is considered as the slack bus. The other nodes in $\mathcal{N}$ represent branch nodes. 
 
Denote a line in $\mathcal{E}$ by the pair ($i,j$) of nodes it connects, where $j$ is closer to the feeder 0. We call $j$ the parent of $i$, denote by $\varsigma (i)$, and call $i$ the child of $j$. Denote the child set of $j$ as $\delta(j) \coloneqq \left \{ i : \left ( i,j \right )\in \mathcal{E} \right \}$. Thus, a link $\left(i,j\right)$ can be denoted as $\left(i,\varsigma (i)\right)$.

For each line $\left(i,\varsigma (i)\right) \in \mathcal{E}$, let $I_{ij}$ be the complex current flowing from nodes $i$ to $\varsigma (i)$, let $Z_{ij} = R_{ij} + \mathbf{i}X_{ij}$ be the impedance of the edge, and $S_{ij}=P_{ij}+\mathbf{i}Q_{ij}$ be the complex power flowing from nodes $i$ to $\varsigma (i)$. On each node $i \in{\mathcal{N}}$, let $V_{i}$ be the complex voltage, and $S_{i}=P_{i}+\mathbf{i}Q_{i}$ be the net complex power injection. Define $v_i\coloneqq \left | V_{i} \right |^2$. We assume the complex voltage $V_{0}$ at the feeder root node is given and fixed. Let $\mathbf{V}=\left [ \mathbf{v^{1},\ldots,v^{N}} \right ]$ be the concatenation of voltage vectors in all nodes in the network.

 

\subsection{Local energy trading under network constraints}

In this subsection, we illustrate the importance of considering the physical network constraints in the trading models, while Section \ref{methodology} provides the description of our methodology.

\begin{figure}[!t]
	\centering
	\includegraphics[width=3in]{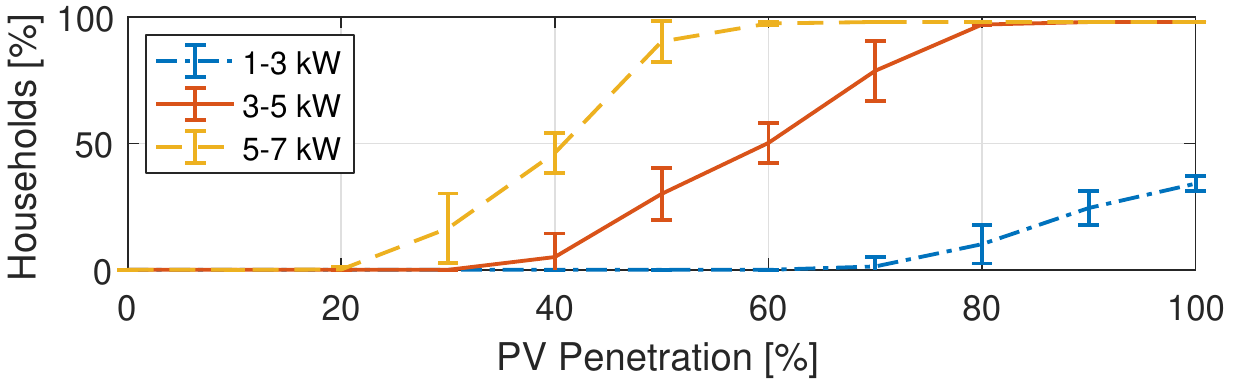}
	\caption{Percentage of households with voltage problems.}
	\label{OV_1}
\end{figure}

\begin{figure}[!t]
	\centering
	\includegraphics[width=3in]{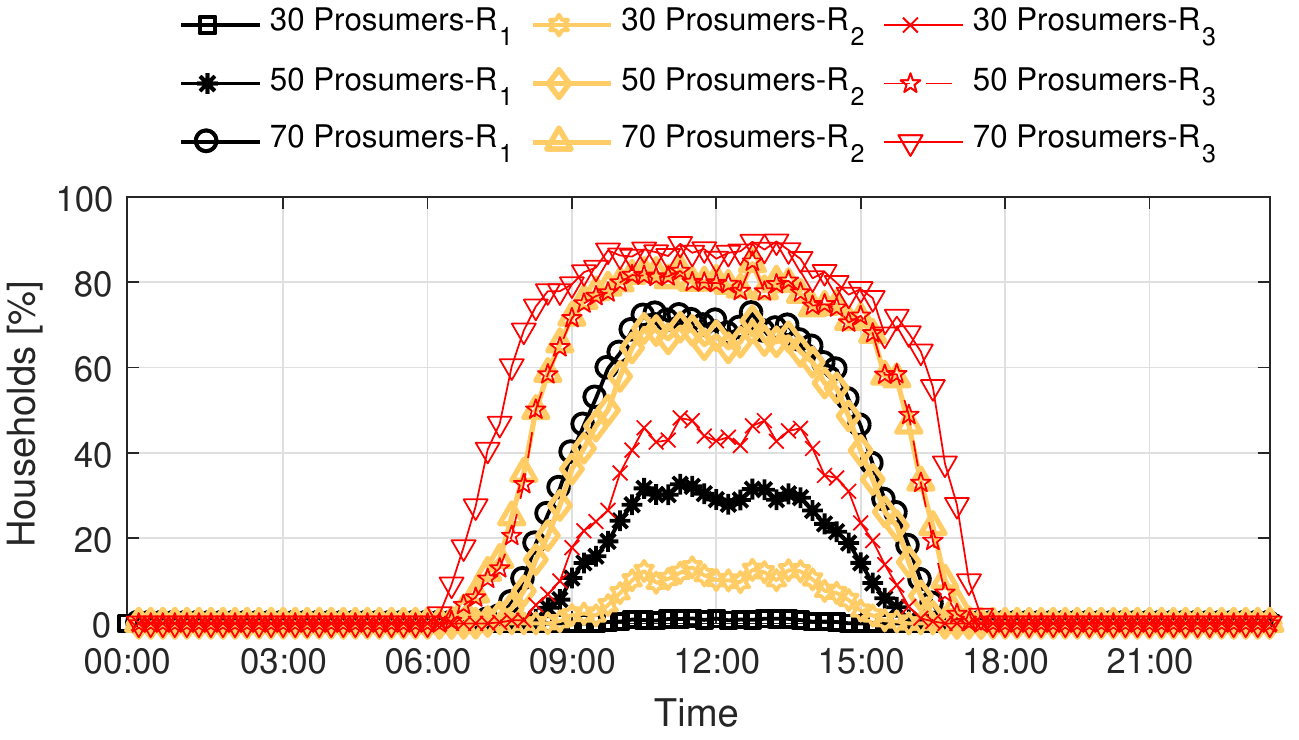}
	\caption{Percentage of households with voltage problems in one day - Under different resistance values and number of prosumers $(R_{1} < R_{2} < R_{3})$.}
	\label{OV_2}
\end{figure}

Many studies in local energy trading have avoided consideration of network constraints to facilitate their modeling\cite{JEGO, ilic_energy_2012, wang_game-theoretic_2014, mengelkamp_trading_2017}. Given that residential users are connected to LV distribution systems, it is necessary to assess the impacts due to the exchange process. Active participation of households without any control could cause network issues such as overvoltage and reverse flows tripping protection equipment \cite{Nando}.

For example, let us assume that a group of end-users in a particular LV network participates in P2P energy trading without considering the network constraints in their trading mechanism, and the energy traded is supplied by non-dispatchable generation such as PV systems. \textcolor{black}{Based on the probabilistic impact assessment methodology proposed in \cite{Nando}, we evaluated the voltage issues at different levels of PV penetrations.} Fig. \ref{OV_1} shows the percentage of households with voltage problems (overvoltage) at different levels of PV penetration. For this feeder, in the worst case, problems start at a penetration of 20\% when the size of the PV systems is between 5 kW and 7 kW. The situation is better with smaller PV sizes. Nevertheless, voltage issues are experienced in all cases. Fig. \ref{OV_2} illustrates the impact of PV penetration using different types of conductors in the network, showing the voltage issues may be worst for networks with greater resistance values. Throughout the day (Fig. \ref{OV_2}), the most critical situation happens around midday (peak of PV generation). Similarly, the situation is worst when there are more prosumers in the network.

Given this context, many strategies have been proposed to prevent the approaching LV network issues. While some methods leave the responsibility to the distribution system operator (DSO; e.g. grid reinforcement, and \textcolor{black}{active transformers with on-load-tap changers \cite{GRID_R,GRID_R2}.}), other strategies consider the direct participation of end-users. For example, PV generation can be curtailed proportionally to avoid voltage problems using a \textit{dynamic curtailment method} \cite{APC}. This method brings benefits to weak nodes which could be highly restricted due to their location in the network, but requires designing a cost-sharing model among prosumers to guarantee fair conditions in the curtailment. In contrast, we show that local energy markets can efficiently allocate the energy surplus, and enable mutual benefits for distribution system operator and all users.

Apart from network issues, technical constraints also influence market efficiency. Since there are external costs associated with power flows, those externalities could represent a barrier to efficient markets. Those extra costs could be internalized in the trading offers of agents. A principled way of addressing the problem of \textit{DER dispatch subject to network constraints} is to use distribution optimal power flow (DOPF)\cite{DLMP}, which is formulated as follows:
\vspace{-0.1in}

\begin{align}
& \underset{P_{i}^{c},P_{i}^{p}}{\text{maximize}} \quad
& & \sum_{i \in \mathcal{N}}C_i^{c}P_i^{c} - \sum_{i \in \mathcal{N}/0}C_i^{p}P_i^{p} \label{ceneq:1} \\
& \text{s.t.} 
& & \text{power flow constraints,} \nonumber \\
&&& \text{power balance constraints, and} \nonumber \\
&&& \text{DER operational constraints,} \nonumber
\end{align}

\noindent where $C_i^{c}$ is the marginal benefit of consumers, $C_i^{p}$ is the marginal cost of prosumers, $P_i^{c}$ is the real power consumption, and $P_i^{p}$ is the real power generation.

DOPF produces distribution locational marginal prices (DLMPs) which can be used to attribute the network cost to the market. In doing so, a central entity (e.g. DSO) solves the optimization problem across the scheduling horizon with the goal of minimizing the total cost of supplying power to the consumers subject to network constraints. As a result, real power losses, and (binding) capacity and voltage constraints result in DLMPs being different across the network. Conceptually, DOPF is the same as the optimal power flow (OPF) problem used in the wholesale market.

The DOPF implementation is however riddled with technical and market design barriers. First, the number of market agents (consumers) is significantly larger than in the conventional OPF problem so that a centralized DOPF computation can be challenging if not intractable. To this end, distributed optimization approaches have been considered \cite{dualdecom_MVC,AA} to ensure scalability. Next, the problem decomposition needs to be done at a household level to preserve consumers’ prerogative and privacy \cite{algorithmic_CVJ,BB}. Finally, household consumption patterns are stochastic, so proper mechanisms are required to ensure that the customers follow the allocated power profiles \cite{faithful_MVC,algorithmic_CVJ,CC}. Second, a DOPF implementation would require a complete redesign of the tariff structures, so it cannot be easily incorporated into the existing market framework.

A viable alternative to the DOPF approach which obviates many of the above challenges is decentralized P2P. However, a successful P2P approach needs to obey the network constraints, as discussed next.



\section{Methodology} \label{methodology}

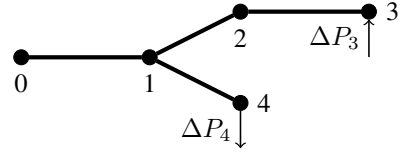
\begin{figure}[t]
	\centering
\begin{tikzpicture}[scale=0.4]
\begin{scope}[node distance=1.5cm]
\node (a1) [ball, label=below:1] at (1,1) {};
\node (a0) [ball, left=of a1, label=below:0] {} edge [ultra thick] (a1);
\node (a2) [ball, above right=of a1, label=below:2, yshift=-10mm] {} edge [ultra thick] (a1);
\node (a3) [ball,right=of a2, label=right:3] {} edge [ultra thick] (a2);
\node (a4) [ball,below right=of a1, label=right:4, yshift=+10mm] {} edge [ultra thick] (a1);
\node (outrow) [ball_2, below=of a4, yshift=+10mm] {} edge [<-,semithick] node [ball_2,left]{$\Delta P_{4}$} (a4);
\node (inrow) [ball_2, below=of a3, yshift=+10mm] {} edge [->,semithick] node [ball_2,left]{$\Delta P_{3}$} (a3);
\end{scope}
\end{tikzpicture}
\caption{A simple distribution network.}
\label{4nodes}
\end{figure}

In this section, we propose a methodology to implement P2P energy trading under network constraints with self-interested agents. This situation is similar to the bilateral trading in a power system. Fig. \ref{4nodes} illustrates the situation where a user located at Bus $4$ has purchased energy from the prosumer located at Bus $3$. This implies physical changes in the power flows through the lines in the network. Hence, our aim is to estimate the impact of the injection and absorption of that amount of power on the grid.

The methodology proposed in this work embeds analytically derived sensitivity coefficients to guarantee bilateral transactions as well as internalizing the external costs associated with the power flows. Specifically, we incorporate three factors in the market mechanism:

\begin{itemize}
	\item[$\bullet$] \textit{Voltage sensitivity coefficients} (VSCs): Through VSCs, we can estimate the variation in the voltages as a function of the power injections in the network;
    \item[$\bullet$] \textit{Power transfer distribution factors} (PTDFs): These reflect the changes in active power line flows due to an exchange of active power between two nodes; 
    \item[$\bullet$] \textit{Loss sensitivity factors} (LSFs): These reflect the portion of system losses due to power injections in the network. 
\end{itemize}

\vspace{-0.1in}

\subsection{Voltage Sensitivity Coefficients Formulation}

The traditional approach to obtain the VSCs is to use the Jacobian matrix after solving the Newton-Raphson power flow \cite{JACOB}:

\begin{equation}
\begin{aligned}
J=\begin{bmatrix}
 \frac{\partial P}{\partial \left | V \right |}&\frac{\partial P}{\partial \theta } \\ 
 \frac{\partial Q}{\partial \left | V \right |}&\frac{\partial Q}{\partial \theta } 
\end{bmatrix},
\end{aligned}
\end{equation}

\noindent where $P$ and $Q$ are the vectors of real and reactive nodal injections, and $\theta$ and $V$ are the vectors of voltage angles and magnitudes. Calculating the inverse of the Jacobian at a given operating point gives an idea of the voltage changes ($\Delta V_{i}$) due to changes in power injections ($\Delta P_{i}$, $\Delta Q_{i}$) as follows:

\begin{equation}
\begin{aligned}
\Delta V_{i}=\left ( \frac{\partial V_{i}}{\partial P_{i}} \right )\Delta P_{i} + \left ( \frac{\partial V_{i}}{\partial Q_{i}} \right )\Delta Q_{i}.
\end{aligned}
\end{equation}

\noindent However, running a full load power flow every time the state of the network changes may not be feasible or tractable. Therefore, in our study, we use the analytical derivation of VSCs proposed in \cite{SENS_1}. In doing so, we use the so-called compound admittance matrix. The relation of the power injection and bus voltages is given by\footnote{Complex conjugate numbers are denoted with a star above (e.g. $V^{*}$)}:
\vspace{-0.1in}
\begin{equation}
\begin{aligned}
S_{i}^{*}=V_{i}^{*}\sum_{j\in{\mathcal{N}}}Y_{ij}V_{j}  \;    i\in{\mathcal{N}}.
\end{aligned}
\end{equation}

\noindent To obtain VSCs, the partial derivatives of the voltages with respect to the active power $P_{k}$ of a Bus $k \in \mathcal{N}/0$ are computed. The partial derivatives with respect to active power satisfy the following system of equations:

\begin{equation} \label{derivatives}
\begin{aligned}
\mathbbm{1}_{\left \{ i=k \right \}}=\frac{\partial V_{i}^{*}}{\partial P_{k}}\sum_{j\in{\mathcal{N}}}Y_{ij}V_{j} + V_{i}^{*}\sum_{j\in{\mathcal{N}/0}}Y_{ij}\frac{\partial V_{j}}{\partial P_{k}}.
\end{aligned}
\end{equation}

\noindent Although this system is not linear over complex components, it is linear with respect to $\frac{{\partial V_{i}}}{{\partial P_{k}}}$ and $\frac{{\partial V^{*}_{i}}}{{\partial P_{k}}}$, therefore it is linear over real numbers with respect to rectangular coordinates. Moreover, it has a unique solution, and can be used to compute the partial derivatives. Once they are obtained, the partial derivatives of the voltage magnitude are expressed as:

\begin{equation} \label{Mag_Sens}
\begin{aligned}
\frac{{\partial \left | V_{i} \right |}}{{\partial P_{k}}}=\frac{1}{\left | V_{i} \right |}\text{Re}\left ( V_{i}^{*}\frac{{\partial V_{i}}}{{\partial P_{k}}} \right ),
\end{aligned}
\end{equation}

\begin{equation} \label{Mag_Sens_2}
\begin{aligned}
{\Delta\left | V_{i} \right |}=\frac{\Delta P_{k}}{\left | V_{i} \right |}\text{Re}\left ( V_{i}^{*}\frac{{\partial V_{i}}}{{\partial P_{k}}} \right ).
\end{aligned}
\end{equation}

\noindent Voltage changes can therefore be calculated based on the power changes in specific buses of the network. 
\vspace{-0.05in}
\subsection{Power Transfer Distribution Factors}

Since the exchange of energy involves power flow through physical routes, PTDFs can give an idea of the sensitivity of the active power flow with respect to various variables. Specifically, the injection shift factor (ISF) quantifies the redistribution of power through each branch following a change in generation or load on a particular bus. It reflects the sensitivity of a flow through a branch with respect to changes in generation or load. Once we obtain the ISFs, we can calculate the PTDFs, which capture the variation in the power flows with respect to the injection in Bus $i$ and a withdrawal of the same amount at Bus $j$  \cite{PTDF,powerbook}. 

In order to calculate the ISFs, we use the reduced nodal susceptance matrix. The ISF of a branch $(k,l) \in \mathcal{E}$ (assume positive real power flow from Bus $k$ to $l$ measured at Bus $k$) with respect to Bus $i \in \mathcal{N}$, which we denote by $\Psi_{kl}^{i}$, is the linear approximation of the sensitivity of the active power flow in branch $(k,l)$ with respect to the active power injection at Bus $i$ with the location of the slack bus specified and all other quantities constant.  Suppose $P_{i}$ varies by a small amount, $\Delta P_{i}$, and let $\Delta P_{kl}^{i}$ be the change in the active power flow in branch $(k,l)$ (measured at Bus $k$) resulting from $\Delta P_{i}$. Then, it follows that:
\vspace{-0.05in}
\begin{equation}
\begin{aligned}
\Psi_{kl}^{i} \coloneqq \frac{\partial P_{kl}}{\partial P_{i}}\approx \frac{\Delta P_{kl}^{i}}{\Delta P_{i}}.
\end{aligned}
\end{equation}

\noindent To calculate these values, we use an approximation of the network equations. Let $\tilde{B}=\text{diag}\left \{ b_{kl} \right \}$, which is a diagonal matrix whose entries are $b_{kl}$, the susceptance of branch $(k,l)$. Also, denote the branch-to-node incidence matrix by $A=\left [ ...,a_{kl},... \right ]^{'}$, where $a_{kl} \in \mathbb{R}^{n}$ is a vector in which the $k^{\text{th}}$ entry is 1 and the $l^{\text{th}}$ entry is -1. Then, by using the DC approximations, we arrive at the expression:

\begin{equation}
\begin{aligned}
\Delta P_{kl}\approx \tilde{B}_{kl}AB^{-1}\Delta P,\\
\end{aligned}
\end{equation}

\noindent where $\tilde{B}_{kl}$ is the row in $\tilde{B}$ corresponding to branch $(k,l)$, and $B=A^{'}\tilde{B}A$. Denote $\Psi_{kl}=\left [ \Psi_{kl}^{1},...,\Psi_{kl}^{i},...,\Psi_{kl}^{N} \right ]^{'}$, then the model-based linear sensitivity factors for branch $(k,l)$ with respect to active power injections at all buses are given by:

\begin{equation} \label{ISFs}
\begin{aligned}
\Psi_{kl} =\tilde{B}_{kl}AB^{-1}.
\end{aligned}
\end{equation}

Once the ISFs are obtained, we compute PTDFs. A PTDF,  $\Phi_{kl}^{ij}$, provides the sensitivity of the active power flow in branch $(k,l)$ with respect to an active power transfer of a given amount of power, $\Delta P_{ij}$, from Bus $i$ to $j$. The PTDF for a branch $(k,l)$ with respect to an injection at a Bus $i$ that is withdrawn at a Bus $j$ is calculated directly from the ISFs as follows:
\begin{equation} \label{PTDFs}
\begin{aligned}
\Phi_{kl}^{ij}=\Psi_{kl}^{i}-\Psi_{kl}^{j},
\end{aligned}
\end{equation}

\noindent where $\Psi_{kl}^{i}$ and $\Psi_{kl}^{j}$ are the line flow sensitivities in branch $(k,l)$ with respect to injections at Buses $i$ and $j$, respectively. 

\subsection{Loss Sensitivity Factors}

We derived the LSFs using a similar approach to the use above. The term for the LSF is given by \cite{Z_BUSS}:

\begin{equation} \label{PlossPk}
\begin{aligned}
\frac{\partial P_{\text{loss}}}{\partial P_{k}}=2\text{Re}\left [  \textbf{V}^{*T}G\frac{\partial \mathbf{V}}{\partial P_{k}} \right ],
\end{aligned}
\end{equation}

\noindent where the partial derivatives are obtained from (\ref{derivatives}), and $G$ is the conductance matrix. In order to assign losses associated to a changes in the power, we consider the approach to attribute losses to bilateral exchanges. For example, in the bilateral exchange in Fig. \ref{4nodes}, there is a bilateral exchange from Bus 3 to Bus 4. The terms $\frac{\partial P_{\text{loss}}}{\partial P_{i}}$ and $\frac{\partial P_{\text{loss}}}{\partial P_{j}}$ are  the loss sensitivities with respect to power injection at bus ${i}$ and to power out at Bus $j$ respectively. Then the \textit{bilateral exchange coefficient} (BEC) is defined as follows:

\begin{equation} \label{BEC}
\begin{aligned}
\text{BEC}^{ij}=\frac{\partial P_{\text{loss}}}{\partial P_{i}}-\frac{\partial P_{\text{loss}}}{\partial P_{j}}.
\end{aligned}
\end{equation}

\noindent The bilateral exchange coefficient (BEC) can be used to associate the losses due to a bilateral transaction \cite{BilateralE}.

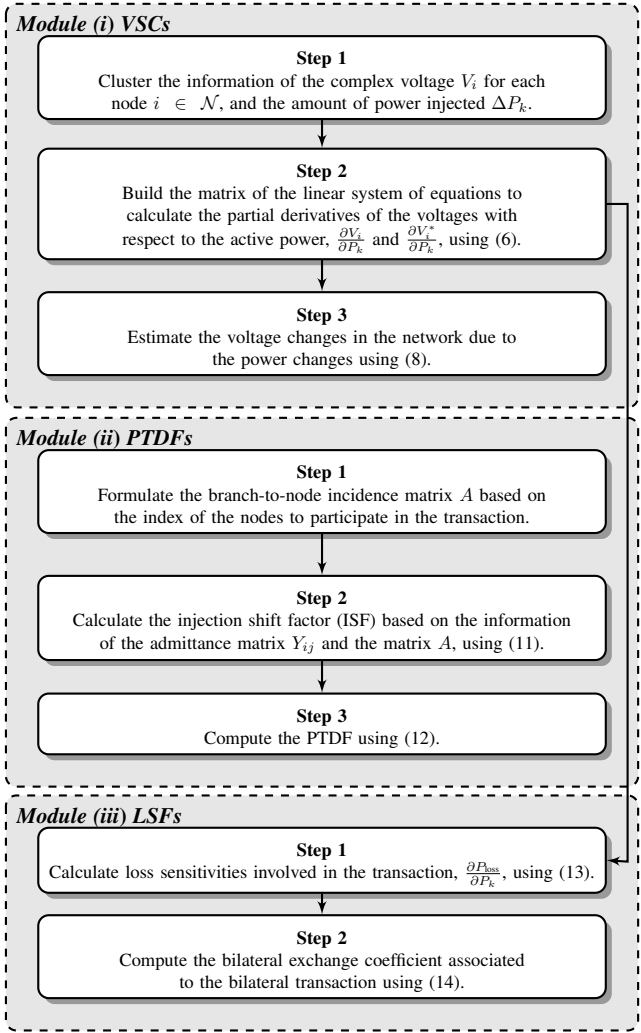
\begin{figure}[t]
\begin{center}
\begin{tikzpicture}[scale=0.7,transform shape]
\path \practica {1}{\textbf{Step 1}\\ Cluster the information of the complex voltage $V_{i}$ for each node $i \in \mathcal{N}$, and the amount of power injected $\Delta P_{k}$.};
\path (p1.south)+(0.0,-1.6)\practica{2}{\textbf{Step 2}\\ Build the matrix of the linear system of equations to calculate the partial derivatives of the voltages with respect to the active power, $\frac{{\partial V_{i}}}{{\partial P_{k}}}$ and $\frac{{\partial V^{*}_{i}}}{{\partial P_{k}}}$, using (\ref{derivatives}).};
\path (p2.south)+(0.0,-1.4) \practica{3}{\textbf{Step 3}\\ Estimate the voltage changes in the network due to \\the power changes using (\ref{Mag_Sens_2}).};

\path (p3.south)+(0.0,-2.2) \practica{4}{\textbf{Step 1}\\Formulate the branch-to-node incidence matrix $A$ based on \\the index of the nodes to participate in the transaction.};
\path (p4.south)+(0.0,-1.6) \practica{5}{\textbf{Step 2}\\Calculate the injection shift factor (ISF) based on the information \\of the admittance matrix ${Y_{ij}}$ and the matrix $A$, using (\ref{ISFs}).};
\path (p5.south)+(0.0,-1.2) \practica{6}{\textbf{Step 3}\\Compute the PTDF using (\ref{PTDFs}). };

\path (p6.south)+(0.0,-2.0) \practica{7}{\textbf{Step 1}\\ Calculate loss sensitivities involved in the transaction, $\frac{\partial P_{\text{loss}}}{\partial P_{k}}$, using (\ref{PlossPk}).};
\path (p7.south)+(0.0,-1.2) \practica{8}{\textbf{Step 2}\\Compute the bilateral exchange coefficient associated \\to the bilateral transaction using (\ref{BEC}).};

\path [line] (p1.south) -- node [above] {} (p2);
\path [line] (p2.south) -- node [above] {} (p3);
    
     \path [line] (p2.east) -- +(0.4,0.0) -- +(0.4,-12.5) -- node [right] {} (p7.east);
\path [line] (p4.south) -- node [above] {} (p5);
    
\path [line] (p5.south) -- node [above] {} (p6);
\path [line] (p7.south) -- node [above] {} (p8);

\background{p1}{p1}{p3}{p3}{\begin {enumerate*}  [label=(\roman*\upshape)]
\item VSCs \end {enumerate*} }
\background{p4}{p4}{p4}{p6}{\begin {enumerate*} [start=2,label=(\roman*\upshape)] \item PTDFs \end {enumerate*} }
\background{p1}{p7}{p8}{p8}{\begin {enumerate*} [start=3,label=(\roman*\upshape)] \item LSFs \end {enumerate*} }

\end{tikzpicture}
    \caption[]{Overview of the methodology. Modules to calculate: \begin {enumerate*}  [label=(\roman*\upshape)]
\item VSCs, \item PTDFs, and \item LSFs. \end {enumerate*}   }
    \label{flowchartNEW}
\end{center}
\end{figure}

An overview of the methodology is shown in Fig. \ref{flowchartNEW}.

\subsection{Illustrative example}

We present a simple example case to illustrate how a bilateral transaction is associated with real power losses, congestion and voltage constraints. We consider a simple five node model shown in Fig. \ref{4nodes}, and we apply the methodology explained in this Section. We assume that the prosumer at Node 3 wants to exchange energy with the consumer at Node 4. That is, an amount of power injected at Node 3 ($\Delta P_{3}$) and is withdrawn at Node 4 ($\Delta P_{4}$). From this transaction, we can obtain the following parameters:

\begin{itemize}
	\item[$\bullet$] Voltage variations caused by the transaction can be estimated using VSCs and (\ref{Mag_Sens_2}). The transaction will not be allowed if   it causes voltage issues in the network. 
	\item[$\bullet$] The PTDFs values, ($\Phi_{01}^{34},\Phi_{12}^{34},\Phi_{23}^{34},\Phi_{14}^{34}$), are calculated to evaluate the utilization rate of the lines based on the transaction. These values can be used to assign congestion charges. As such, agents will pay a charge for using the physical network. Moreover, PTDFs can be used to estimate the congestion in the lines.
	\item[$\bullet$] The total system losses caused by the transaction are calculated using the VSCs and (\ref{PlossPk}) and (\ref{BEC}). Therefore, agents involved in the transaction will be responsible for paying an extra cost due to the losses caused using coefficient $\text{BEC}^{34}$.
\end{itemize}

These elements allow us to evaluate the impact of each transaction in the network, and they can be used to incorporate more properties to the model. For example, since users will have to pay the extra cost due to congestion and losses, users will tend to prefer to exchange energy with the closest ones. 

\section{Trading Market Mechanism}

The market mechanism for a P2P energy trading developed in this paper builds on our previous work \cite{JEGO}. There are three components to our market mechanism: \begin {enumerate*} [label=(\roman*\upshape)]
\item a \textit{continuous double auction} (CDA), \item the agents' bidding strategies, and \item the network permission structure, as described below.
\end {enumerate*} 
\subsection{Continuous Double Auction}

A CDA matches buyers and sellers in order to allocate a commodity. It is widely used, including in major stock markets like the NYSE. A CDA is a simple market format that matches parties interested in trading, rather than holding any of the traded commodity itself. This makes it very well suited for P2P exchanges. Bids into a CDA indicate the prices that participants are willing to accept a trade, and reflect their desire to improve their welfare. As such, the CDA tends towards a highly efficient allocation of commodities \cite{gode_allocative_1993}. In more detail, a CDA comprises:

\begin{itemize}
	\item[$\bullet$] A set of \textit{buyers} $\mathcal{B}$, where each $b\in{\mathcal{B}}$ defines its trading price $\pi_{b}$ and the amount of energy to purchase $\sigma_{b}$.
	\item[$\bullet$] A set of \textit{sellers} $\mathcal{S}$, where each $s\in{\mathcal{S}}$ defines its trading price $\pi_{s}$ and the amount of energy to sell $\sigma_{s}$.
	\item[$\bullet$] An \textit{order book}, with bids $o_{b}(b,\pi_{b},\sigma_{b},t)$, made by buyers $\mathcal{B}$, and asks $o_{s}(s,\pi_{s},\sigma_{s},t)$, made by sellers $\mathcal{S}$.	
\end{itemize}

\noindent \textcolor{black}{Pseudo-code of the matching process in a CDA is given in Algorithm \ref{zi}. A CDA is run for each time slot separately. Any intertemporal couplings that arise on a customer's side from using batteries or loads with long minimum operating times are not passed up to the market clearing entity}. Once the market is open, arriving orders are queued in the \textit{order book} for trades during a fixed interval $t_{d}$ (lines 2-8)\textcolor{black}{, which is limited by the start time $t_{d}^{\text{st}}$ and the trading end time $t_{d}^{\text{end}}$ (i.e. $t_{d}^{\textmd{end}} = t_{d}^{st} + t_{d}$). During the trading period, orders are submitted for buying or selling units of electrical energy in time-slot t. At the end of the trading period, the market closes, thereby no more offers are received.”} 
We assume the orders arrive according to a Poisson process with mean arrival rate $\lambda$. The current best bid (ask) is the earliest bid (ask) with the highest (lowest) price. 
\textcolor{black}{A bid and an ask are matched when the price of a new bid (ask) is higher than or equal to the price of the best ask $o_{s}^{*}(s^{*},\pi_{s}^{*},\sigma_{s}^{*},t^{*})$ (the best bid $o_{b}^{*}(b^{*},\pi_{b}^{*},\sigma_{b}^{*},t^{*})$) in the order book (line 9)}.
However, if a new bid (ask) is not matched, then it is added to the order book, recording its arrival time and price. Note that after matching, an order may be only partially covered. If this is the case, it will remain at the top of the order book waiting for a new order. This process is executed continually during the trading period as new asks and bids arrive.

\subsection{Bidding Strategies}

Conventionally, market participants (buyers and sellers) define their asks and bids based on their preferences and the associated costs. The HEMS act as agents for the customers, and are continually responding to new stochastic information. As such, they appear very unpredictable from the outside. Moreover, because the market is thin, this can produce large swings in available energy and prices. In this context, constructing an optimal bidding strategy is futile, but simple bidding heuristics are still valuable. In particular, in our study the agents are \textit{zero intelligence plus} (ZIP) traders \cite{ZIP,JEGO}. ZIP traders use an adaptive mechanism which can give performance very similar to that of human traders in stock markets. \textcolor{black}{Agents have a profit margin which determines the difference between their limit prices and their asks or bids. Under this strategy, traders adapt and update their margins based on the matching of previous orders (lines 12-23 for buyers and lines 24-35 for sellers).} Indeed, the participation of ZIP traders in a CDA allows us to assess the economic benefits of the market separate from that of a particular bidding strategy. Specifically, ZIP traders are subject to a budget constraint ($L_{\text{max}}$ and $L_{\text{min}}$ are the maximum and minimum price respectively) which forbids the trader to buy or sell at a loss. Then, buyers and sellers select their bids or asks uniformly at random between these limits.

\begin{algorithm}[t]
	\footnotesize
	\caption{Matching process in a CDA with ZIP traders}
	\label{zi}
	\begin{algorithmic}[1]
		\While {market is open}
		\State randomly select a new ZIP trader
		\If{buyer} 
		\State	new $o_{b}(b,\pi_{b},\sigma_{b},t)$ 
		\Else 
		\State	new $o_{s}(s,\pi_{s},\sigma{s},t)$ 
		\EndIf
		\State 	allocate a new order in the \textit{order book}
		\LeftComment{Evaluate matching process with best bid and ask}
		\If{$\pi_{b}^{*} \geq \pi_{s}^{*}$}
		\State clear orders $o_{b}^{*}$ and $o_{s}^{*}$ at a price $\pi_{t}$ and amount $\sigma_{t}$
		\EndIf
		\LeftComment{Update values of profit margins}
		\LeftComment{Buyers} 
		\If{the last order was matched at price $\pi_{t}$} 
		\State	all buyers for which $\pi_{b}\geq \pi_{t}$, raise their margins;  
		\If {the last trader was a seller}
		\State any active buyer for which $\pi_{b}\leq \pi_{t}$, 
		\State lower its margin;
		\EndIf
		\Else 
		\If {the last trader was a buyer}
		\State any active buyer for which $\pi_{b}\leq \pi_{t}$, 
		\State lower its margin;
		\EndIf			
		\EndIf
		\LeftComment{Sellers} 
		\If{the last order was matched at price $\pi_{t}$} 
		\State	all sellers for which $\pi_{s}\leq \pi_{t}$, raise their margins;  
		\If {the last trader was a buyer}
		\State any active seller for which $\pi_{s}\geq \pi_{t}$, 
		\State lower its margin;
		\EndIf
		\Else 
		\If {the last trader was a seller}
		\State any active seller for which $\pi_{s}\geq \pi_{t}$, 
		\State lower its margin;
		\EndIf			
		\EndIf
		
		\EndWhile		
	\end{algorithmic}
\end{algorithm}

\subsection{Network Permission Structure}

\begin{figure}[t]
	\centering
	\begin{tikzpicture}[scale=0.1,auto, node distance=3cm,>=latex']
    \node [process, name=input, scale=0.8] {\small Received continually asks \& bids};
    \node [qtion, above of=input, scale=0.6, yshift=-5em] (q1) {\large Matched?};
    \node [block, right of=q1, scale=0.8,xshift=-2em] (p1) {\small Estimate voltage and power flow variations};
    \node [block, right of=p1, scale=0.8,xshift=-2em] (p2) {\small Update network state estimation. Block high risk households.};
     \node [below of=p1, yshift=+3.6em, scale=0.8, yshift=+3em] (users) {Households};
     \node [block2, below of=users, yshift=+7em, scale=0.8] (prosumers) {\small Prosumers};
     \node [block2, below of=prosumers, yshift=+7em, scale=0.8] (consumers) {\small Consumers};
    
    \node [process2, right of=p2, scale=0.8,xshift=-4em] (output) {\small Allocate extra costs};
    
    \node [draw, below of=consumers, scale=0.8, yshift=+7.5em] (open) {\small Open};

     \node [above of=users, yshift=-8.1em] (housetop) {};
     \node [left of=prosumers, xshift=+6.5em,, yshift=-1.5em] (houseleft) {};
     \node [right of=houseleft, xshift=-4.3em] (houseright) {};

        \draw [draw,->] (input) -- node {} (q1);
     \draw [->] (q1.east) -- node [near start]{\footnotesize Yes} (p1);
     \draw [->] (p1.east) -- node [name=y] {}(p2);
     \draw [->] (houseleft) -- node [above] {\footnotesize $o_{s},o_{b}$}(input.east);
    \draw [->] (p2.east) -- node [name=y] {}(output);

	\draw [->] (p2.south) |- node [name=y] {\footnotesize $\phi^{1}, \phi^{2},\ldots, \phi^{H}$}(houseright);
    \draw [->] (q1.west) |- node [near start] {\footnotesize No}(input.west);
    \draw [draw,->] (open.west) -| node {} (input.south);
    
    \begin{pgfonlayer}{background}
        \path (q1.west |- p2.north)+(-0.5,0.3) node (a) {};
        \path (input.south -| p2.east)+(+0.5,-0.2) node (b) {};
        \path[fill=gray!20,rounded corners, draw=black!50, dashed]
            (a) rectangle (b);
        \path (users.north west)+(-0.4,0.1) node (a) {};
         \path (consumers.south east)+(+0.3,-1.3) node (b) {};
        \path[fill=blue!10,rounded corners, draw=black!50]
             (a) rectangle (b);
    \end{pgfonlayer}
\end{tikzpicture}
\caption{Schematic of the P2P trading under network constraints.}
\label{BD}
\end{figure}
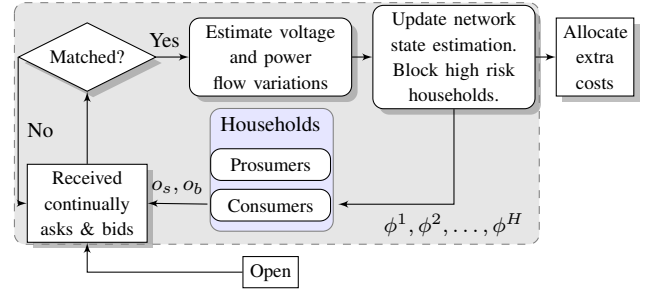

The outline of the mechanism is presented in Fig. \ref{BD}. \textcolor{black}{A third party entity (e.g. DSO) validates the transactions using a network permission structure based on the network's features and sensitivity coefficients}. Every time one ask and one bid are matched, voltage variation and line congestion are evaluated. All households receive a signal ($\phi^{h}$) which informs them if they can still participate in the market without causing problems in the network. For instance, one prosumer could be blocked from injecting power into the grid at a certain time due to the high risk of causing voltage problems in the network. This is achieved using the VSCs and PTDFs. If the transaction is approved, the extra cost associated with the network constraints are allocated to the users involved in the matched transaction. 

Importantly, power curtailment is implicitly incorporated in the trading. Thus, this method may bring extra benefits in comparison to others curtailment methods. For example, users at the worst node location still have the opportunity to participate if their order can be matched and if the mechanism allows the trade. This improves the efficiency by allowing greater participation of consumers and a better reflect of network conditions. 

\section{System Model - Case Study}
Our study is focused on a LV network with a high DERs penetration. The group of households is constituted by consumers and prosumers (Type 1 and Type 2) defined in Section II. There are three components to our model: the local power \textit{network}, the \textit{customers} and the \textit{market} for trading energy, as defined above.
\vspace{-0.01in}
\subsection{Implementation: Test Network}
\begin{figure}[t]
	\centering
	\includegraphics[scale=1]{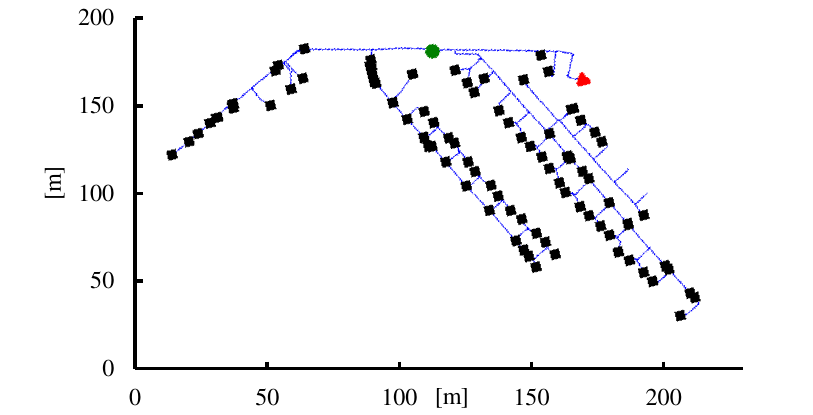}
	\caption{Topology of the studied LV network. The black squares, the green point and the red triangle in the topology, represent the location of households, the CES, and the transformer, respectively.}
	\label{network}
\end{figure}

We consider a smart grid system for energy trading at a local level. The methodology is applied to the UK LV network shown in Fig. \ref{network}, comprising one feeder and 100 single phase households. The simulations are carried out with $T=24$ hours, $\Delta \tau = 15$ minutes and up to 100 agents. There are 50 consumers and 50 prosumers, 40 for Type 1 (PV) and 10 for Type 2 (PV, battery and HEMS). Each household has a stochastic load consumption profile, with load profiles using the tool presented in \cite{CREST}. Similarly, PV profiles are generated considering sun irradiance data, capturing the sunniest days in order to evaluate the method on the most challenging yet realistic scenarios. We assume that all prosumers have a PV system with installed capacity of 5.0 kWp. Each Type 2 households has a battery of 3 kW and 10 kWh. 

Additionally, there is one \textit{community electricity storage} (CES) of 25 kW and 50 kWh operated by the retailer. In particular, the operation objective of the CES is to apply peak shaving during peak load hours. The CES’ strategy is to buy only the energy to charge in the P2P market to other prosumers around midday (when there are low rates and a high number of prosumers with energy surplus) and resell the energy during peak demand hours to the consumers. Like the prosumers behavior, the CES is modeled as a ZIP trader.

We define the price constraints $L_{\text{max}}$ and $L_{\text{min}}$ based on the values of import and export electricity tariffs through the day. $L_{\text{max}}$ depends on the time-of-use tariff (ToU) and $L_{\text{min}}$ on the feed-in-tariff (FiT). These definitions are consistent in the sense that no buyer would pay more than the tariff of a retailer (ToU), and no seller would sell their units cheaper than the export tariff (FiT). In summary, the process of our model is:

\begin{enumerate}
	\item The HEMS minimizes a prosumer's costs by solving problem (\ref{HEMS}), using a mixed-integer linear program.
	\item Prosumers state the time-slots when they have extra energy to trade.
	\item The bidding strategies for the market participants are initialized, using their load and generation profiles and tariffs, and the market is opened.
    \item Every time an ask and a bid are matched, the network conditions are evaluated. The market remains open as long as the network constraints are respected.
	\item Agents accept the number of units to be exchanged and their prices.
\end{enumerate}
\vspace{-0.01in}
\subsection{Scenarios' Description}

Since our interest is to evaluate our methodology and to show the benefits of P2P energy trading under network constraints, two scenarios are evaluated.
\subsubsection{Scenario I}
The first scenario is based on the methodology introduced in this paper. Users participate in P2P trading. The matching process between asks and bids in the P2P market promotes the local balance of demand and generation of end-users. In this case, a market rule allows the prosumers to supply their energy surplus until the total demand, including the energy required by the CES, is covered. 
\subsubsection{Scenario II}
In this case, prosumers are allowed to inject more energy into the grid as long as that does not cause any voltage or capacity problems in the network. Since curtailment methods are commonly used to prevent LV network issues in a high PV penetration, we considered them as a benchmark in this scenario. As such, we compared our scheme with other curtailment methods to illustrate the benefits of the local markets and the extra benefits of power curtailment functionality. Specifically, the four schemes to compare are:

\begin{itemize}
	\item[$\bullet$] \textit{Local market P2P} (P2P): The methodology introduced in this paper.
	\item[$\bullet$] \textit{Reduce capacity} (Red. Cap): A static active power curtailment method. All users can export only a limited power to the grid. In this case, all prosumers can export $\leq$ 3 kW. This value is chosen based on an impact assessment study of this particular network. It ensures the network constraints are not violated.
	\item[$\bullet$] \textit{Tripping}: The standard approach where an inverter operates until it reaches the maximum voltage limit. Then, the inverter protection shuts it down.	
    \item[$\bullet$] \textit{Droop-based active curtailment} (APC-OLP): A dynamic active power curtailment method. Inverters are controlled with a droop-based active power curtailment method (APC). The droop parameters of the inverters are different so that the output power losses (OPL) are shared equally among all prosumers \cite{APC}.
\end{itemize} 

For the three benchmark schemes, households buy energy at the ToU rate and sell at the FiT value. \textcolor{black}{Each scheme is simulated using OpenDSS software. We consider a daily simulation mode using the same input data for all schemes. The operation settings of PV systems is modified depending on the features of each scheme (e.g. 3 kW is the maximum power to export to the grid in Red.Cap case).}


\vspace{-0.1in}
\subsection{Scenario I Results}
\begin{figure}[!t]
	\centering
	\includegraphics[scale=0.69]{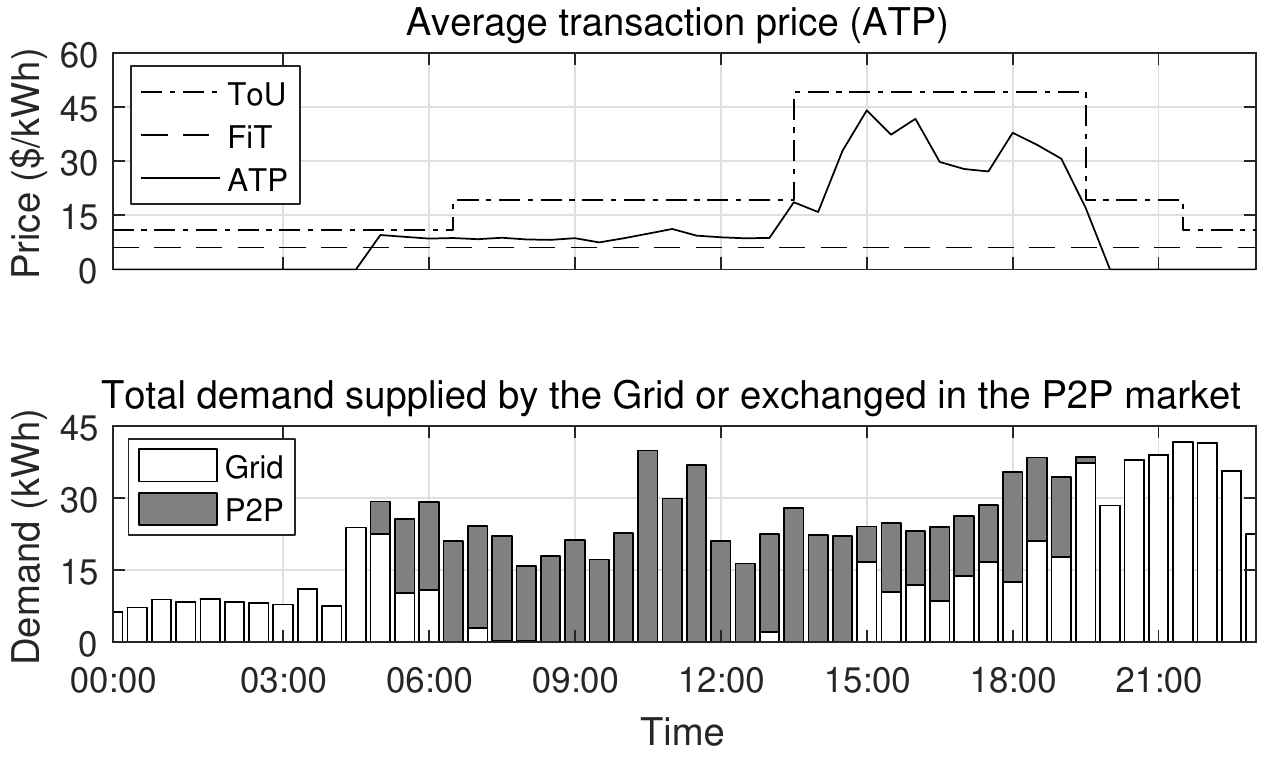}
	\caption{Average transaction prices (top), demand and generation levels (bottom) in Scenario I.}
	\label{BalancedM}
\end{figure}

Fig. \ref{BalancedM} shows the average transaction price (ATP) and the amount of energy purchased from the grid or in the P2P market during one day. The transaction prices remain in the range of ToU and FiT rates because of the ZIP limits $L_{\text{max}}$ and $L_{\text{min}}$. Hence, both prosumers and consumers obtain monetary benefits by participating in P2P trading.
Most of the energy is traded during 8:00 and 14:00. During that time, there is an excess of energy due to PV generation. Notably, there is a peak of energy sold in the market around 11 am because of the charging strategy of the CES. 
There is some energy traded after 18:00 due to the CES and the prosumers who kept some energy in the battery. Once the peak time ends (20:00), the ZIP maximum limit $(L_{\text{max}})$ is low. As a consequence, no prosumers submit any new asks to trade in the market. 
Moreover, in this case, when the total energy surplus from prosumers is greater than the total demand of consumers (e.g. around midday), some prosumers (those who do not match their asks with consumers' bids) have to curtail their power generation. 

\begin{figure}[!t]
	\centering
	\includegraphics[scale=0.66]{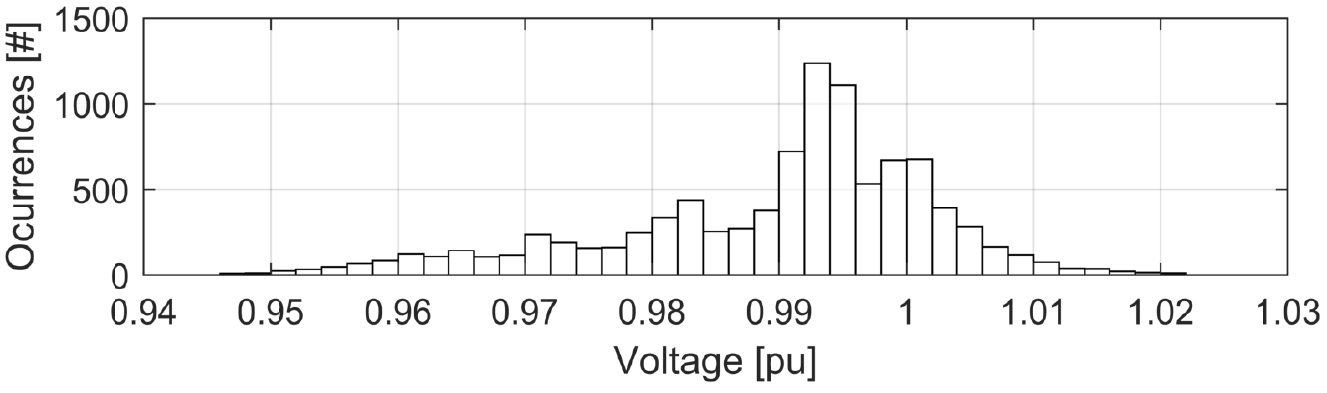}
	\caption{Histogram of voltages at users' nodes - number of occurrences in one day period at a certain voltage [pu] in Scenario I.}
	\label{voltageH}
\end{figure}

Fig. \ref{voltageH} presents a histogram of voltages at all users' nodes during one day of simulation. There are no cases of overvoltage. The voltages varied between 0.945 pu and 1.022 pu. Around 55\% of the voltages are between 0.99 pu and 1 pu. As such, all exchanges respect the network constraints, and the external costs were attributed among the households involved in each transaction. 

Finally, Table \ref{table_1} compares the total expenses and incomes of all households during one day. Without the P2P trading, end-users buy energy at the ToU rate and sell it at the FiT value. In contrast, with P2P, the transaction prices are discovered through the market mechanism. Hence, the users' expenses decrease and the users' incomes increase, achieving a market benefit of \$75.92, while remaining within the networks operating limits.

\begin{table}[!t]
\centering
\caption{Comparison of total expenses and incomes in Scenario I}
\label{table_1}
\begin{tabular}{|cc|cc|c|}
\hline
\multicolumn{2}{|c|}{\textbf{Without P2P}} & \multicolumn{2}{c|}{\textbf{With P2P}} & \multirow{2}{*}{\textbf{\begin{tabular}[c]{@{}c@{}}Market \\ Benefit\end{tabular}}} \\ \cline{1-4}
Expenses             & Incomes             & Expenses           & Incomes           &                                                                                     \\ \hline
\$241.98             & \$32.37             & \$198.50           & \$64.81           & \$75.92                                                                             \\ \hline
\end{tabular}
\end{table}
\vspace{-0.1in}
\subsection{Scenario II Results}

\begin{figure}[t]
	\centering
	\includegraphics[scale=0.68]{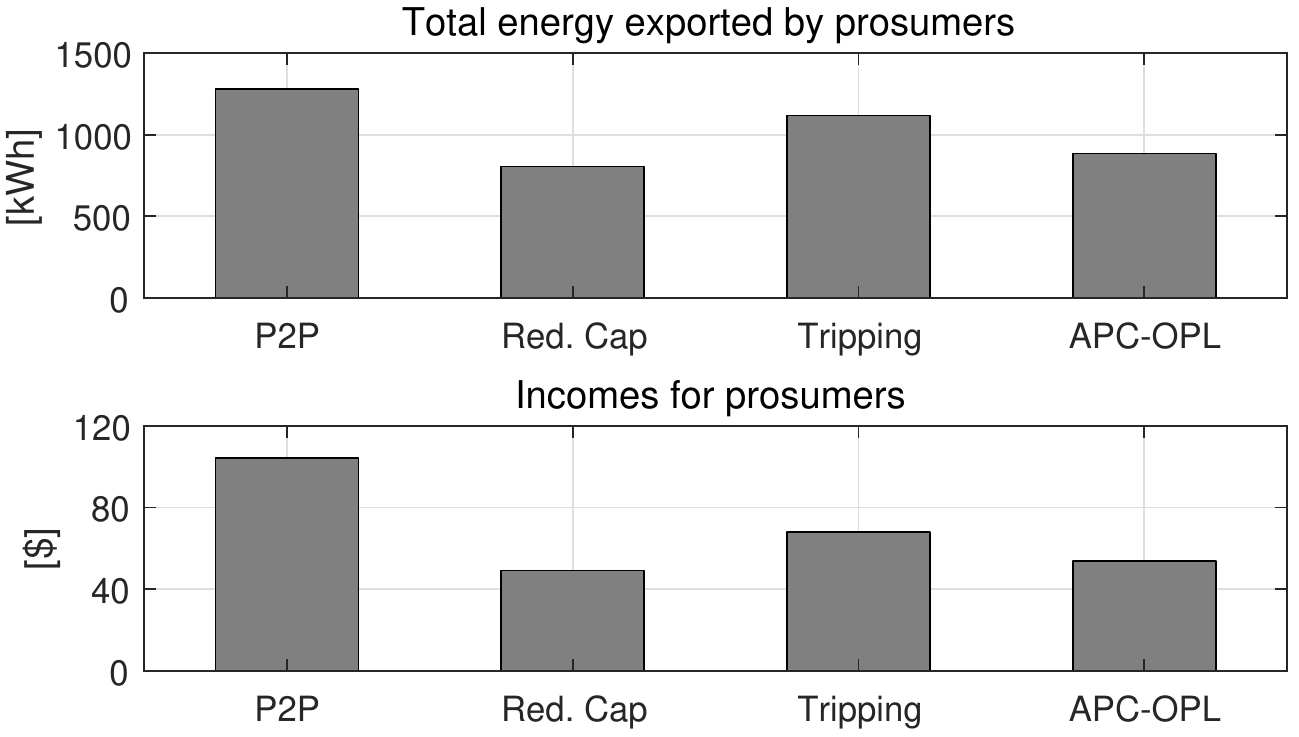}
	\caption{Total energy supplied to the grid by prosumers and their incomes received in Scenario II.}
	\label{comparison}
\end{figure}

This scenario compares our method with the benchmark curtailment schemes. The results in Fig. \ref{comparison} show that in the P2P case there is more energy traded, and the revenues for the prosumers are greater in comparison with the other methods. Hence, this local market reduces the energy spilled and increases the prosumers' incomes. Particularly, the drawback of the power curtailment methods is that they do not consider the impact on the revenues of end-users. In contrast, the P2P scheme offers greater economic benefits to all users. For example, in the \textit{Tripping} case, the furthest prosumer (with respect to the location of the feeder) is regularly the first to be curtailed, and its energy spilled is 70\% of its total energy surplus. In contrast, the energy spilled is only around 50\% in the P2P case. So, the prosumer sold more energy in the P2P case, thereby its income increased by \$0.7. In this way, the P2P local market provides distributed coordination, control and management of the DERs. 

\vspace{-0.05in}
\section{Conclusion}
In this paper, we have proposed a new methodology to deploy P2P energy trading local markets considering the network constraints in the market mechanism. We explicitly considered the impact of the injection and absorption of power in the network in a P2P exchange. Users exchange energy with their neighbors through a continuous double auction, and their transaction internalized the extra cost associated with the technical constraints. Simulation results showed that our proposed method reduces the energy cost of the users and achieves the local balance between generation and demand of households without violating the technical constraints. Finally, we compared the implementation of our market with other curtailment methods. Our technique captures the desirable properties of curtailment methods with the market platform. Hence, our system exploits profitable opportunities for reduced spilled energy to all stakeholders. 

\textcolor{black}{Due to the use of a continuous double auction (CDA), the proposed method doesn’t suffer from the scalability issues of OPF and DLMP models. Specifically, stock exchanges allow for huge numbers of trades a day (e.g. NASDAQ processes 10M trades each day). This is actually a key benefit of the CDA approach, because the complexity is kept on the trading agent side of the ledger, not the clearing entity. In a standard CDA, the clearing entity has only very low computation routines to complete. While this P2P framework has an additional bid permission overlay, the complexity of these routines is not great (i.e. no optimization) and the number of bids on a typical MV feeder is not expected to exceed that of a stock exchange.}

\textcolor{black}{The future work will extending the study of bidding strategies of agents with flexible loads participating in a P2P market, as well as the incorporation of penalty policy to evaluate prediction deviations in forecast profiles and to enhance the trading among nearby users. 
}


%





\ifCLASSOPTIONcaptionsoff
  \newpage
\fi



%




\bibliographystyle{IEEEtran}
\bibliography{bibtex/bib/test}

%
\vskip -2\baselineskip plus -1fil
\begin{IEEEbiography}
[{\includegraphics[width=1in,height=1.25in,clip,keepaspectratio]{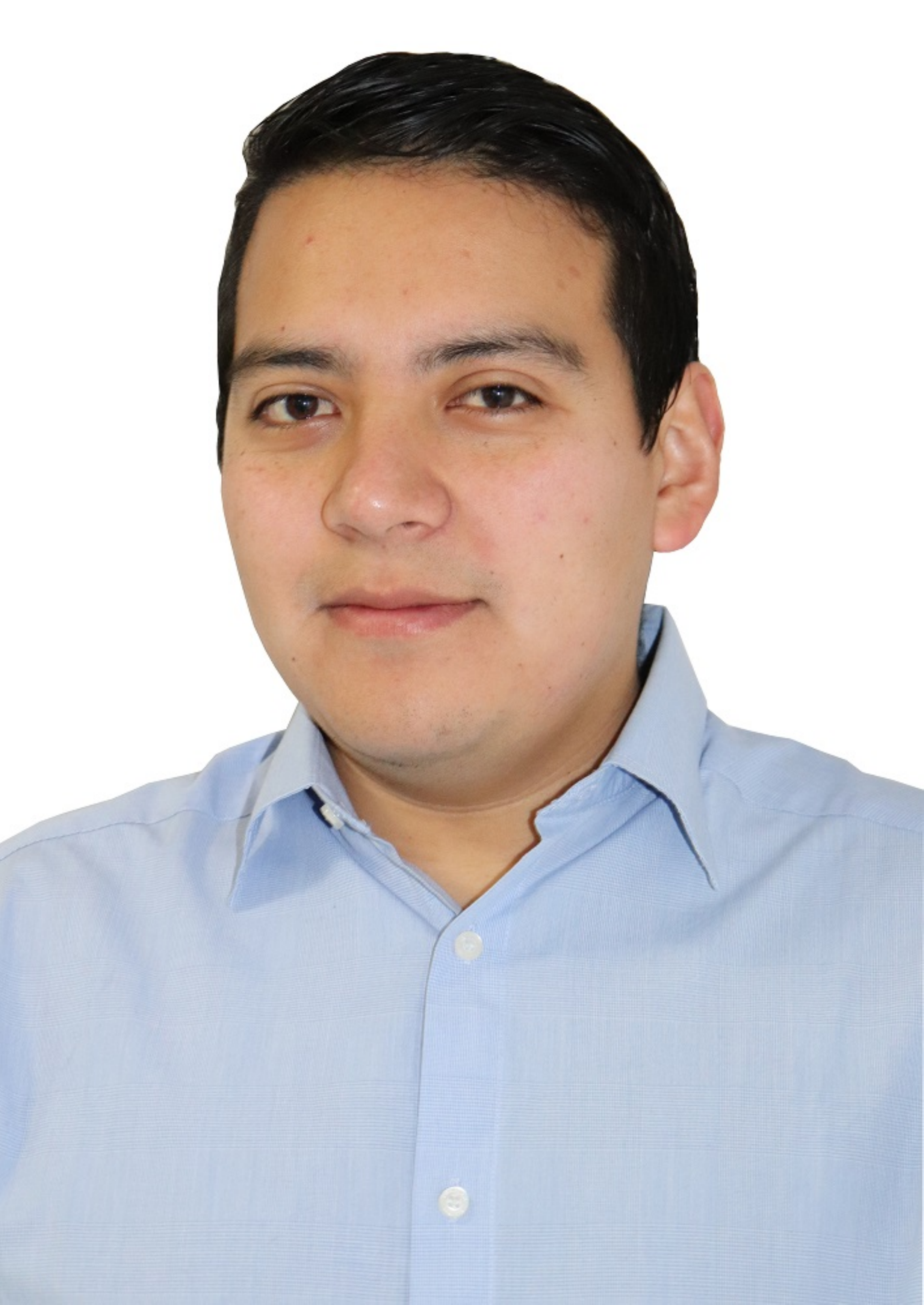}}]
{Jaysson Guerrero}
(S$\text{'}$10) was born in Pasto, Colombia. He received the B.Sc. degree in electronics engineering, B.Sc. degree in electrical engineering, and the M.Sc. degree in electrical engineering from the Universidad de los Andes, Bogot\'{a}, Colombia, in 2013, and 2014, respectively. He is currently pursuing the Ph.D. degree in Electrical Engineering at The University of Sydney. His research interests include integration of renewable energy into power systems, smart grid technologies and local energy trading.
\end{IEEEbiography}
\vskip -2\baselineskip plus -1fil
\begin{IEEEbiography}
[{\includegraphics[width=1in,height=1.25in,clip,keepaspectratio]{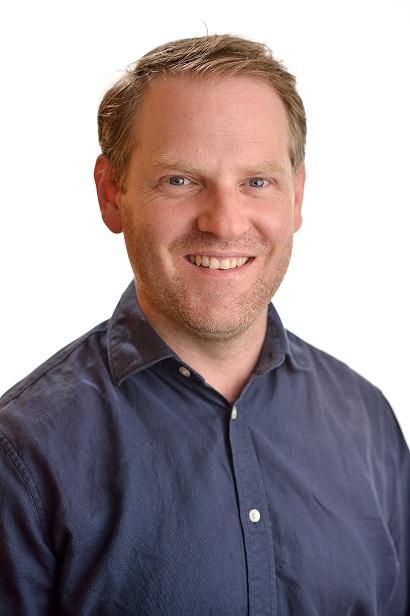}}]
{Archie C. Chapman}
(M$\text{'}$14) received the B.A. degree in math and political science, and the B.Econ. (Hons.) degree from the University of Queensland, Brisbane, QLD, Australia, in 2003 and 2004, respectively, and the Ph.D. degree in computer science from the University of Southampton, Southampton, U.K., in 2009. 
He is currently a Research Fellow in Smart Grids with the School of Electrical and Information Engineering, Centre for Future Energy Networks, University of Sydney, Sydney, NSW, Australia. 
His work focuses on the use of distributed energy resources, such as batteries and flexible loads, to provide power network and system services, while making best use of legacy infrastructure.  
His expertise is in optimization and control of large distributed systems, using methods from game theory and artificial intelligence.
\end{IEEEbiography}

\vskip -2\baselineskip plus -1fil
\begin{IEEEbiography}[{\includegraphics[width=1in,height=1.25in,clip,keepaspectratio]{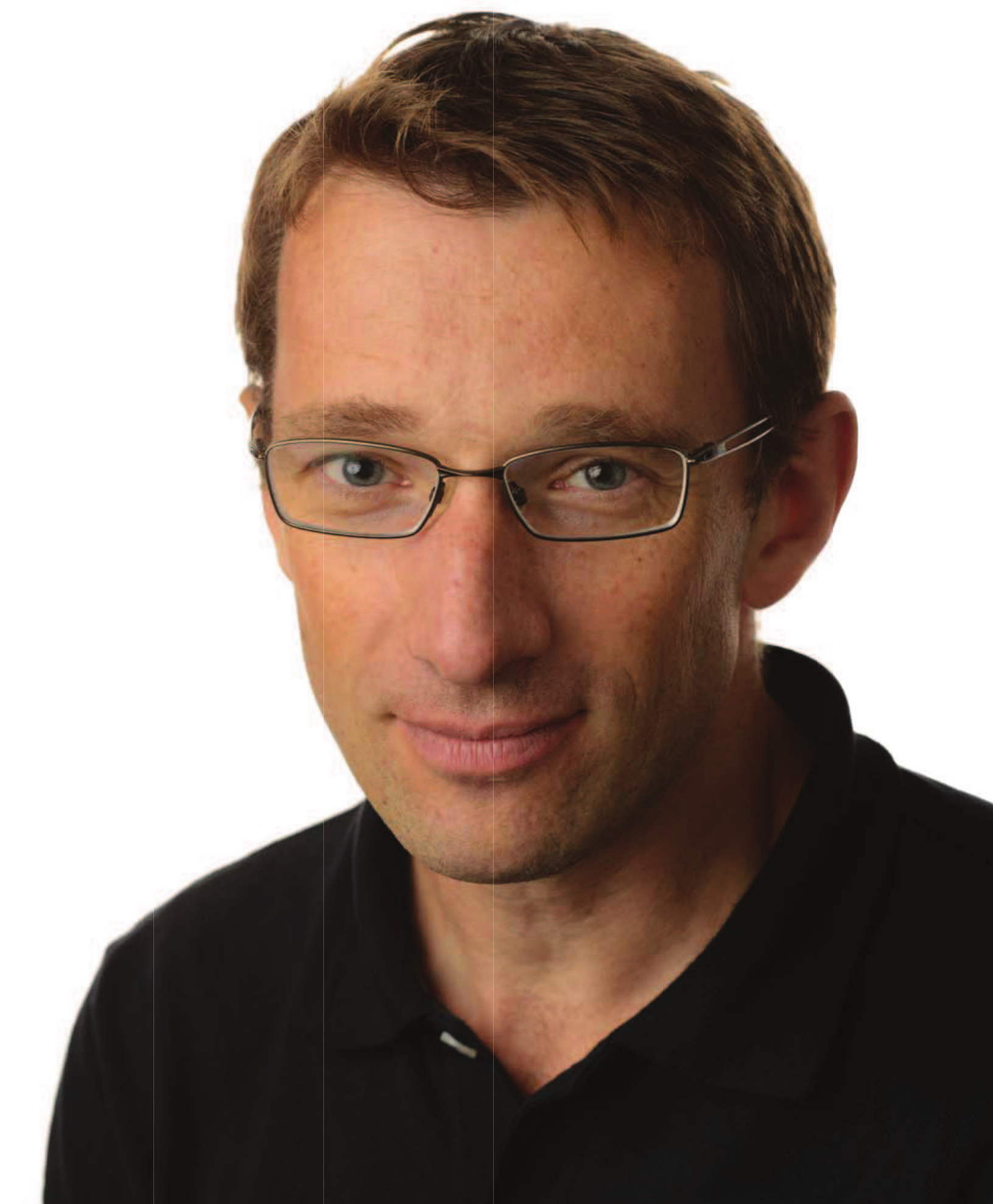}}]{Gregor Verbi\v{c}} (S'98-M’03-SM'10) received the B.Sc., M.Sc., and Ph.D. degrees in electrical engineering from the University of Ljubljana, Ljubljana, Slovenia, in 1995, 2000, and 2003, respectively. In 2005, he was a NATO-NSERC Postdoctoral Fellow with the University of Waterloo, Waterloo, ON, Canada. Since 2010, he has been with the School of Electrical and Information Engineering, The University of Sydney, Sydney, NSW, Australia. His expertise is in power system operation, stability and control, and electricity markets. His current research interests include grid and market integration of renewable energies and distributed energy resources, future grid modelling and scenario analysis, wide-area coordination of distributed energy resources, and demand response. He was a recipient of the IEEE Power and Energy Society Prize Paper Award in 2006. He is an Associate Editor of the IEEE Transactions on Smart Grid.
\end{IEEEbiography}




\end{document}